%% file: sample-base.tex
\documentclass[conference]{IEEEtran}
\IEEEoverridecommandlockouts
\usepackage{cite}
\usepackage{amsmath,amssymb,amsfonts}
\usepackage{booktabs}
\usepackage{multirow}
\usepackage{multicol}
\usepackage{algorithmic}
\usepackage{graphicx}
\usepackage{textcomp}
\usepackage{xcolor}
\def\BibTeX{{\rm B\kern-.05em{\sc i\kern-.025em b}\kern-.08em
    T\kern-.1667em\lower.7ex\hbox{E}\kern-.125emX}}
\begin{document}

\makeatletter
\newcommand{\linebreakand}{%
  \end{@IEEEauthorhalign}
  \hfill\mbox{}\par
  \mbox{}\hfill\begin{@IEEEauthorhalign}
}
\makeatother

\title{Social Context-aware GCN for Video Character Search via Scene-prior Enhancement
\thanks{
* Tong Xu is the corresponding author.

This work was supported by the grants from National Natural Science Foundation of China (No.U22B2059, 62222213, 62072423), the USTC Research Funds of the Double First-Class Initiative (No.YD2150002009), and the Alibaba Group through Alibaba Innovative Research (AIR) Program.}
}

% \author{
% \IEEEauthorblockN{name1,name2,name3,name4,name5}
% \IEEEauthorblockA{
% Beijing, China \\
% }\\

% \author{
% \IEEEauthorblockN{Wenjun Peng$^1$,\quad Weidong He$^1$,\quad Derong Xu$^2$,\quad Tong Xu$^2$,\quad Chen Zhu$^{3,4}$,\quad Enhong Chen$^2$}
% \IEEEauthorblockA{
% \textit{School of Computer Science and Technology, University of Science and Technology of China
% } 
% }
% }

\author{
\IEEEauthorblockN{Wenjun Peng$^{1,2}$,\quad Weidong He$^{1,2}$,\quad Derong Xu$^{1,2}$,\quad Tong Xu$^{1,2*}$,\quad Chen Zhu$^{1,3}$,\quad Enhong Chen$^{1,2}$}
\IEEEauthorblockA{
\textit{$^1$University of Science and Technology of China,
}\\
\textit{$^2$State Key Laboratory of Cognitive Intelligence,
}\\
\textit{$^3$Alibaba.
}\\
$\{$pengwj,hwd,derongxu$\}$@mail.ustc.edu.cn, $\{$tongxu,cheneh$\}$@ustc.edu.cn, \\
zc3930155@gmail.com.
}
}

% \author{\IEEEauthorblockN{Wenjun Peng}
% \IEEEauthorblockA{\textit{School of Computer Science and Technology} \\
% \textit{University of Science and Technology of China}\\
% Hefei, China \\
% pengwj@mail.ustc.edu.cn}
% \and
% \IEEEauthorblockN{Weidong He}
% \IEEEauthorblockA{\textit{School of Computer Science and Technology} \\
% \textit{University of Science and Technology of China}\\
% Hefei, China \\
% hwd@mail.ustc.edu.cn}
% \linebreakand
% \IEEEauthorblockN{Derong Xu}
% \IEEEauthorblockA{\textit{School of Data Science} \\
% \textit{University of Science and Technology of China}\\
% Hefei, China \\
% derongxu@mail.ustc.edu.cn}
% \and
% \IEEEauthorblockN{Tong Xu}
% \IEEEauthorblockA{\textit{School of Data Science} \\
% \textit{University of Science and Technology of China}\\
% Hefei, China \\
% tongxu@ustc.edu.cn}
% \linebreakand
% \IEEEauthorblockN{Chen Zhu}
% \IEEEauthorblockA{\textit{School of Management} \\
% \textit{University of Science and Technology of China}\\
% Hefei, China \\
% zc3930155@gmail.com}
% \and
% \IEEEauthorblockN{Enhong Chen}
% \IEEEauthorblockA{\textit{School of Data Science} \\
% \textit{University of Science and Technology of China}\\
% Hefei, China \\
% cheneh@ustc.edu.cn}
% }

\maketitle

\begin{abstract}
With the increasing demand for intelligent services of online video platforms, video character search task has attracted wide attention to support downstream applications like fine-grained retrieval and summarization. However, traditional solutions only focus on visual or coarse-grained social information and thus cannot perform well when facing complex scenes, such as changing camera view or character posture. Along this line, we leverage social information and scene context as prior knowledge to solve the problem of character search in complex scenes. Specifically, we propose a scene-prior-enhanced framework, named SoCoSearch. We first integrate multimodal clues for scene context to estimate the prior probability of social relationships, and then capture characters' co-occurrence to generate an enhanced social context graph. Afterwards, we design a social context-aware GCN framework to achieve feature passing between characters to obtain robust representation for the character search task. Extensive experiments have validated the effectiveness of SoCoSearch in various metrics.
\end{abstract}


\input{sections/intro}
\input{sections/relatedworks}
\input{sections/framework}

\input{sections/experiments}

\input{sections/conlusion}

% \section{Acknowledgement}
% This work was supported by the grants from National Natural Science Foundation of China (No.U22B2059, 62222213, 62072423), the USTC Research Funds of the Double First-Class Initiative (No.YD2150002009), and the Alibaba Group through Alibaba Innovative Research (AIR) Program.
\bibliographystyle{IEEEbib}
\bibliography{sample-base}

\end{document}

%% file: sections/intro.tex
\section{INTRODUCTION}
As an indispensable technology for supporting various video applications, video character search plays an important role in downstream tasks such as clip retrieval, video summarization, etc. For example, many video platforms, such as Tencent and YouKu Video, provide an ``only look at him" function by searching the clips of specific characters in videos to meet fans' needs, which enhances user experience and brings significant business benefits. However, the manual collection of clips for specific characters requires a lot of human resources.

\begin{figure}[t]
	\centering
	\includegraphics[width=1\linewidth]{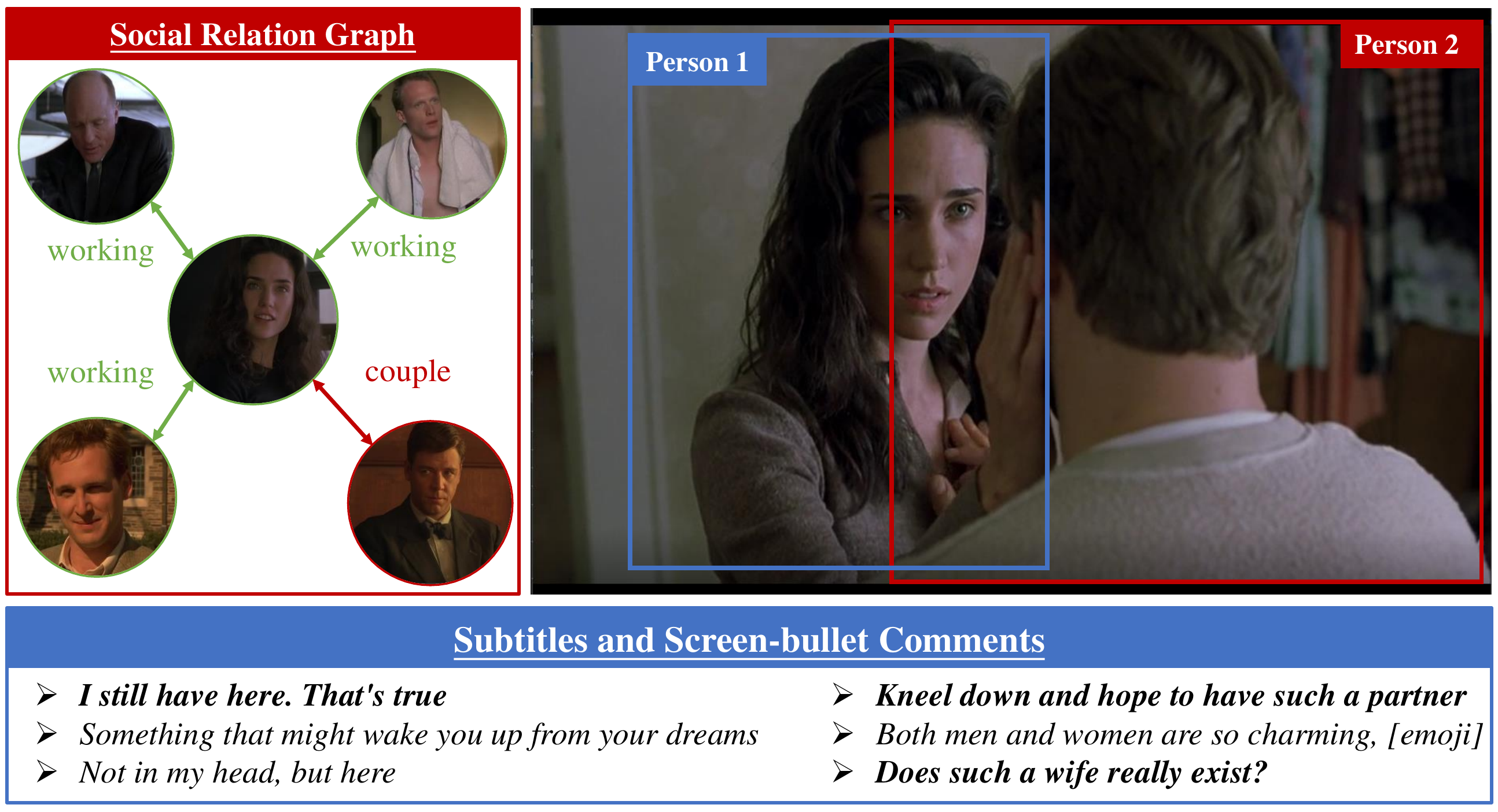}
	\caption{A film scene and its social relation graph.}
	\label{inferidentity}
    \vspace{-0.4cm}
\end{figure}

To this end, automatic character search attaches much concentration from academia and industry. Nevertheless, most mainstream video character search methods mainly focus on visual information or coarse-grained social information to identify characters. 
For example, classic video character search methods, such as~\cite{zhou2019omni, wu2019where}, only focus on extracting visual clues and ignoring other meta information. ~\cite{li2021social} tries to leverage social information, but they only use heuristic rules based on co-occurrence and fail to fully use the scene's context to extract the prior probability of relationships, thus limiting their effectiveness. Considering that, with the plot processing, the characters may develop many social relationships. Estimating the impact of these relationships on co-occurrence behavior without discrimination will seriously weaken the differentiation of relationships. Let us take Figure~\ref{inferidentity} as an example. There are two characters in this film scene, where the female is easy to identify because of her clear visual clues. But as for the male one, on the one hand, his back cannot provide enough visual clues to recognize him. On the other hand, the numerous relationships of the female also make it difficult to reveal its co-occurrence character. However, with carefully checking the fine-grained information, such as the subtitles and the kinds of social relationships between characters, we can reasonably infer the identity of this male. Obviously, the subtitles and visual clues of the background imply two characters are a ``couple". And the most likely person, dating with the female, is her husband.

Based on the above observations, in this paper, we propose a scene prior knowledge based \textbf{So}cial \textbf{Co}ntext-aware character \textbf{Search} framework named \textbf{SoCoSearch}, which can integrate fine-grained social information and scene context to search video characters effectively. Specifically, SoCoSearch first generates a prior probability for different relationship types to obtain a weighted social graph through integrated multimodal semantic clues. At the same time, we use the characters' visual features to select an anchor node and mine characters' co-occurrence. The co-occurrence serves as an extra clue to enhance the weighted social graph into a social context graph, which helps to learn the candidate set of potential matching characters. Furthermore, we employ a social context-aware GCN to achieve feature passing between graph nodes and obtain characters' social features. Finally, we use anchor node similarity to adjust the weight of visual and social features adaptively. The technical contributions of this article can be summarized as follows:

\begin{itemize}
	\item We further discussed the assistance of social relationships on the video character search task, and took the lead in using the scene context as prior knowledge to enhance the model's performance on the task.
	% 结合社交关系应对任务
	\item We proposed a novel neural-based framework, which effectively integrates multiple clues and automatically adjusts the weight of various features.
	\item The experiments on real video datasets prove that our method outperforms several state-of-the-art baselines in terms of mAP, mINP and R1.
    % 自适应社交关系和视觉权重
\end{itemize}

%% file: sections/relatedworks.tex
\section{RELATED WORKS}
\subsection{Person Search.} The main challenge of traditional methods is extracting robust and discriminative visual features. \cite{yu2017devil, chang2018multi, zhou2019omni} extract multiple level features from earlier layers, and fuse them as the final representation to enhance the performance of models. To guarantee that the model has better generalization ability across domains, \cite{zhuang2021camera} introduces normalization modules by learning domain styles to improve the model's performance across different cameras. \cite{he2021transreid} designs two novel modules named jigsaw patch module and side information embeddings, which can improve the robustness and discrimination ability of the model by rearranging patches and adding domain information.

Nevertheless, traditional datasets are often gathered from a few nearby cameras in a short time, leading to the appearance of the same identity often similar, with slight variance in posture, clothing or occlusion. 
To solve the problem, \cite{yan2019learning} utilizes social context to make models adaptable to the video character search task in unconstrained environments. \cite{li2021social} takes a step further and uses high-level social relations to identify characters. 
Unfortunately, this method relies on heuristic modeling and fails to reveal the prior probability of the relationship in different scenes, thus limiting the guidance capability of characters' occurrence.

\subsection{Multimodal Learning.} Recently, some studies have applied multimodal learning techniques to video applications. \cite{zhou2019character} proposes a novel framework for character-oriented video summarization by combining visual and textual information for the first time. \cite{xu2021socializing} introduces a multi-stream architecture to recognize relations in videos. \cite{kukleva2020learning} proposes a model that considers visual and dialogue clues to recognize interactions and social relations jointly. Inspired by the above studies, we design a dynamic relation weight module to predict the prior probability of relationships in different scenes via multimodal clues.

%% file: sections/framework.tex
\section{TECHNICAL FRAMEWORK}

\begin{figure*}[h]
	\centering
	\includegraphics[width=\linewidth]{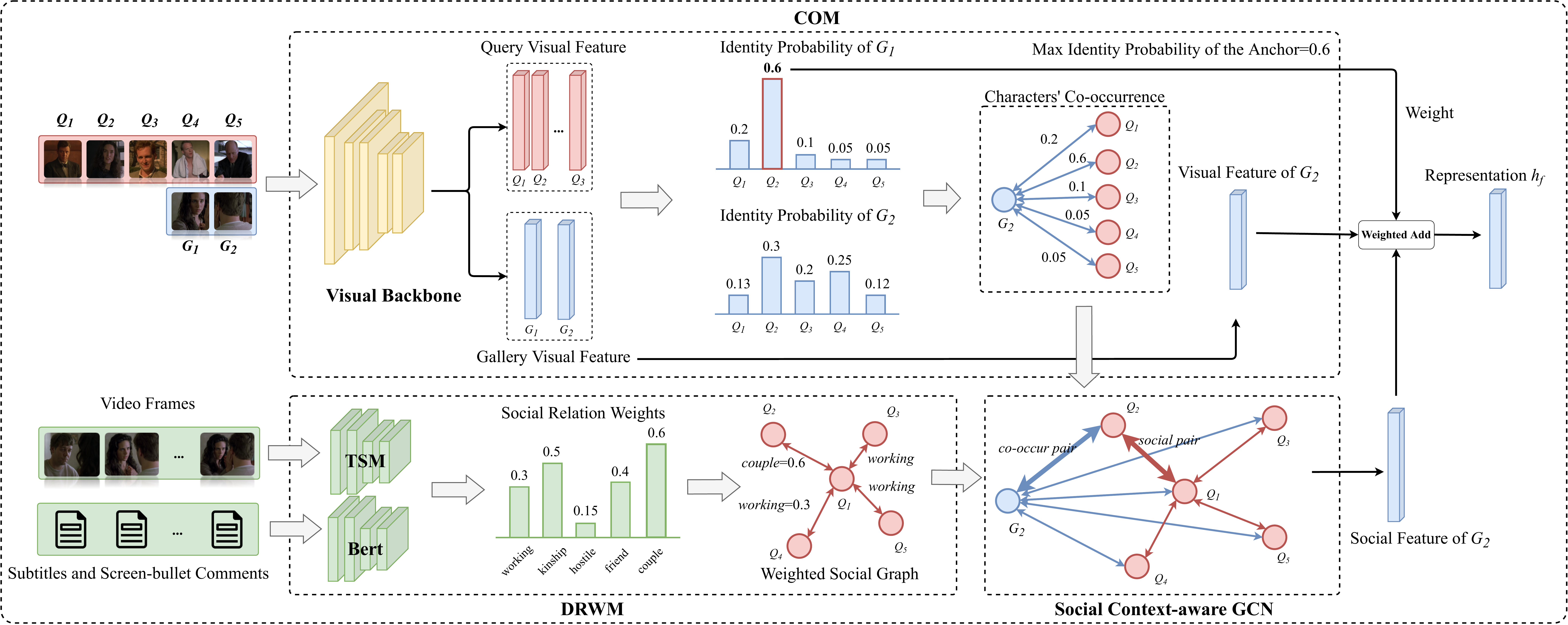}
	\caption{The overall structure of our framework.}
	\label{framework}
    \vspace{-0.4cm}
\end{figure*}

\subsection{Framework Overview}
The video character search task aims to retrieve all gallery RoIs belonging to a specific query character in a social context-aware condition. 
In order to better address this task, as illustrated in Figure 2, we propose a framework that includes a dynamic relation weight module (DRWM), a co-occurrence miner (COM), and a Social Context-aware Graph Convolutional Network (GCN).
Firstly, DRWM generates the prior probability for different social relationships to get a weighted social relation graph. At the same time, COM uses visual features to strengthen the social relation graph. 
Finally, the social features of gallery characters are learned by social context-aware GCN, and integrated with visual features to get the final representation.
%\old{Finally, the anchor similarity will be used to adjust the weight of visual and social feature to get the final representation.}

\subsection{Dynamic Relation Weight Module}
% We design a Dynamic Relation Weight Module(DRWM) to construct a weighted social relation graph. Given the visual context and textual context extracted from time window $T_{t_i}$, DRWM can predict the probability of different relationships in the scene. Then the probability of the relationship will be assigned to the corresponding edges of the predefined social relation graph to get a weighted social relation graph.\par
% hwd
Given the visual and textual context as scene prior knowledge, the Dynamic Relation Weight Module (DRWM) aims to predict the probabilities of different relationships between the characters in the scene. These probabilities are assigned to the corresponding edges of the predefined social relation graph to construct a weighted social relation graph.

% Specifically, this module consists of two main components, which are TSM\cite{lin2019tsm} for vision information and Bert\cite{devlin2018bert} for text information. Suppose the frames and text collected from $T_{t_i}$ are $F_{t_i} = (f_{t_i-m}, f_{t_i-m+1}, ..., f_{t_i+m})$ and $D_{t_i} = (d_{t_i-m}, d_{t_i-m+1}, ..., d_{t_i+m} )$, where $m$ is the window size. We denote the output of TSM, $ h_v = TSM(F_{t_i}) $, as visual feature. And we denote the pooler output of Bert, $h_d= Bert(D_{t_i})$, as the textual feature. \par
% hwd
Specifically, we first leverage TSM\cite{lin2019tsm} and Bert\cite{devlin2018bert} to extract visual and textual features, respectively. $h_v$ is denoted as the visual features, and $h_d$ is denoted as the textual features.
We concatenate $h_v$ with $h_d$, and feed them into a linear layer to obtain the relation weight $\hat{y_r}$. We assign the relation weight to the corresponding edge in the predefined social graph to get the weighted social graph. For the example shown in Figure \ref{framework} (DRWM Part), the relationship between $q_1$ and $q_2$ is $``couple"$ and the corresponding relation weight output from DRWM is 0.6, so the edge weight of $e_{ij}(``couple")$ is assigned to 0.6. 
We formulate the relation recognition as a multi-label classification task, and the binary cross-entropy loss $\mathcal{L_R}$ is expressed as:

\begin{equation}
	\mathcal{L_R} = -y_r \cdot log(\hat{y_r})-(1-y_r) \cdot log(1-\hat{y_r}),
\end{equation}
where $y_r$ is the true relationships between each pair of gallery characters in the scene, and $\hat{y_r}$ is the prediction of the DRWM.

\subsection{Co-occurrence Miner}
Given the RoIs of the query and gallery, Co-occurrence Miner (COM) aims to reveal characters' co-occurrence. In this module, we first calculate the similarity between the RoIs of the query and gallery by euclidean distance. Then, we convert the similarity to the possibility of characters' co-occurrence.

Specifically, we employ TransReID\cite{he2021transreid} as our visual backbone to learn the visual features of characters. 
% Given an image $ x $, TransReID can output its visual feature $h$. 
Afterwards, we calculate the euclidean distance of each query-gallery features pair and then convert it to a co-occurrence probability. The process can be expressed as follows:

\begin{equation}
	d(g_i,q_j)=||\boldsymbol{h_{g_i}} - \boldsymbol{h_{q_j}}||^{2}_2,
    % \vspace{-0.15cm}
\end{equation}

\begin{equation}
	s(g_i,q_j)=\frac{\bar{d}-d(g_i,q_j)}{\delta(d)},
 % \vspace{-0.15cm}
\end{equation}

\begin{equation}
	p(g_i,q_j)=\frac{exp(s(g_i,q_j))}{\sum_{j}{exp(s(g_i, q_j))}},
\end{equation}
where $\boldsymbol{h_{g_i}}$ and $\boldsymbol{h_{q_j}}$ are visual features of gallery $g_i$ and query $q_j$, respectively. $\bar{d}$ is the average distance and $\delta(d)$ is the standard deviation of all distances. $p(g_i, q_j)$ is the identity probability, representing the probability of $g_i$ and $q_j$ belonging to the same character. Then, COM calculates the co-occurrence probability of each pair. For example, in Figure \ref{framework} (COM Part), given five query characters $\{q_{j}| j=1,2,3,4,5\}$, two gallery characters $\{g_{i}|i=1,2\}$ and their respective probability distributions of similarity, $g_1$ is regarded as the anchor character since the highest confidence value 0.6. The distribution of $p(g_1, q_j)$ serves as the possibility of co-occurrence of $p(g_2, q_j)$, which consists of the edge weight of the social context graph. Specifically, $g_1$ and $g_2$ appear simultaneously, and $g_1$ is probably $q_j$. Therefore, it can be considered that $q_j$ is likely to co-occur with $g_2$, and the similarity between $g_1$ and $q_j$ is the co-occurrence probability of $q_j$ and $g_2$. Thus, we connect $g_2$ with all $q_j$ and assign co-occurrence probability to the corresponding edge's weight to represent characters' co-occurrence.
For scenes with more than two characters, we choose the character with the highest similarity as the anchor.

\subsection{Social Context-aware GCN}
% \vspace{-0.1cm}
In order to make full use of characters' relationships and co-occurrence simultaneously, this module combines the weighted social relation graph obtained from DRWM and the characters' co-occurrence obtained from COM to generate the social context graph. Then it performs message passing on the graph to extract social features of the gallery, which are weighted and combined with visual features to learn the final representation.\par
% It is noting that, in this graph, the second-order neighbors of each gallery node consist of candidate character set. For example, as shown in Figure \ref{framework}, on the one hand, $gallery_1$ and $gallery_2$ appear simultaneously in this scene. According to identity probability of gallery, we can find that $gallery_1$ is easy to recognize while $gallery_2$ is hard to identify. So, COM select $gallery_1$ as the anchor node and calculate the similarity between $gallery_1$ and all queries. After that, we know that $gallery_1$ is most likely $query_2$, which means $gallery_2$ is also most likely to appear with $query_2$ among all queries. On the other hand, in the predefined social relation graph, the DRWM assigns weights to different types of edges according to the possibility of different relationships. Furthermore, the relationship most likely to exist is $r$ in this scene. So the nodes that have the relationship $r$ with $query_2$ are likely to be $gallery_2$. Correspondingly, the relationship between $query_1$ and $query_2$ is $r$. To sum up, $gallery_2$ and $query_1$ are most likely to have the same identity. \par
It is noting that, in this graph, the second-order neighbors of each gallery node constitute a candidate character set, which makes the model pay more attention to the characters who are likely to exist in the scene. As mentioned before, DRWM assigns weights to different relationship edges to determine the most likely social pairs in the scene. COM mines the characters' co-occurrence based on visual similarity to determine the most likely co-occurrence pairs in this scene. Finally, as shown in Figure \ref{framework} (Social Context-aware GCN part), we can get meta paths according to social and co-occurrence pairs and obtain the gallery character candidates.
\par

Therefore, in the social co-occurrence graph, we need to aggregate the feature of the second-order neighbors into the corresponding gallery node to reduce the difference between them. In addition, we remove the feature transformation weight $W$ and activation function $f$ to obtain pure features of second-order neighbors. The specific operations are as follows:
\begin{equation}
	\boldsymbol{h^{(l+1)}}=\boldsymbol{\tilde{D}^{-\frac{1}{2}}\tilde{A}\tilde{D}^{-\frac{1}{2}}}\boldsymbol{h^{(l)}},
\end{equation}
where $\tilde{A} \in \mathbb{R}^{N\times N}$ is the adjacency matrix of the graph, and $\tilde{D} \in \mathbb{R}^{N \times 1}$ is the degree matrix of nodes, $h^{(l)} \in \mathbb{R}^{N\times d}$ is the output from $l$ layer. Suppose $h^{(2)}$ is the social feature output from 2-layer social content-aware GCN, $ h^{(0)} $ is the visual feature output from the visual backbone, the final representation of the gallery node is:
\begin{equation}
	w_s = \frac{p_a}{\sum{\hat{y_r}}},
 % \vspace{-0.05cm}
\end{equation}
\begin{equation}
	\boldsymbol{h_{f}} = w_s \boldsymbol{h^{(2)}}+(1-w_s) \boldsymbol{h^{(0)}},
\end{equation}
where $p_a$ donates the highest anchor's identity probability, $\hat{y_r}$  donates the relationships' probability predicted by DRWM, which is used to control the weight of social features. Because when the identity probability of the anchor is very low or there are many relationships in a scene, the possibility that the second-order neighbor of the gallery node is the corresponding query character will also decrease. Therefore, we need to reduce the weight of social feature $h^{(2)}$.\par
% We formulate $\mathcal{L_{ID}}$, $\mathcal{L_T}$ to optimize the COM and the GCN. $\mathcal{L_{ID}}$ is the cross-entropy loss and $\mathcal{L_T}$ is the triplet loss. Suppose $\hat{y}_{id}$, $y_{id}$ are the prediction and groundtruth of the character’s id respectively, $h_f$ is the representation of a gallery character $g_i$, $h_p$ is the positive representation that has same character ID with $g_i$, $h_n$ is the negative representation that has different character ID with $g_i$. The $\mathcal{L_{ID}}$ and $\mathcal{L_T}$ can be expressed as:
% \begin{equation}
% 	\mathcal{L_{ID}} = \sum{-y_{id} \cdot log(\hat{y}_{id})}
% \end{equation}
We formulate the cross entropy loss $\mathcal{L_{ID}}$ and triplet loss $\mathcal{L_T}$ to optimize COM and Social Context-aware GCN, which is expressed as: 
\begin{equation}
    \hat{y}_{id} = \boldsymbol{W_c} * \boldsymbol{h_f} + \boldsymbol{b_c},
    % \vspace{-0.05cm}
\end{equation}
\begin{equation}
	\mathcal{L_{ID}} = \sum{-y_{id} \cdot log(\hat{y}_{id})},
    % \vspace{-0.05cm}
\end{equation}
\begin{equation}
	\mathcal{L_T} = log[1+exp(||\boldsymbol{h_f}-\boldsymbol{h_p}||^{2}_{2}-||\boldsymbol{h_f}-\boldsymbol{h_n}||^{2}_{2})],
\end{equation}
where $W_c$ and $b_c$ is weight and bias of classification layer. $\hat{y}_{id}$, $y_{id}$ are the prediction and groundtruth of the character’s ID respectively. $\boldsymbol{h_f}$ is the representation of a gallery character $g_i$, $\boldsymbol{h_p}$ is the representation of the positive sample that has same character ID with $g_i$, $\boldsymbol{h_n}$ is the representation of the negative sample that has different character ID with $g_i$. Finally, we combine three loss to get the overall loss function:
\begin{equation}
	\mathcal{L = L_R+L_{ID}+L_T}.
\end{equation}

%% file: sections/experiments.tex
\section{EXPERIMENTS}
\subsection{Datasets and Pre-processing}
To the best of our knowledge, there is no off-the-shelf dataset available for the video character search task that contains RoIs of characters and their corresponding video frames, textual information, and relation graph. Therefore, following \cite{li2021social}, we constructed two datasets named Social-Bilibili and Social-MovieNet based on the Bilibili\cite{zhou2019character} and MovieNet\cite{huang2020movienet} datasets, respectively. 

\textbf{Social-Bilibili Dataset}. The dataset contains 70 films with an average length of 2 hours, 375 main characters, related subtitles and screen-bullet comments. All characters' relationships are grouped into five categories, which are working, kinship, hostile, friend and couple respectively. Based on the dataset, we sampled 325 characters, a total of 11478 images, for train, and the remaining 50 characters, 1989 images, for test. Following \cite{li2021social}, we selected a clear image for each character as the query image, which constitutes our query set. In addition, we collect the video frames and textual information with time window $T_t=[t-12, t+12]$ as the context.

\textbf{Social-MovieNet Dataset}. We collected social relation graphs for 16 films from MovieNet, with an average length of 1.7 hours, 79 main characters and related subtitles. To be consistent with the Social-BiliBili dataset, we also group all relationships into five categories, which are working, kinship, hostile, friend and couple respectively. We sampled 59 characters, a total of 14462 images, for train, and the remaining 20 characters, 5991 images, for test. Similarly, we selected a clear image for each character as a query image, which constitutes our query set. Since MovieNet has split movies into shots, we set the time window to $T_t=[t-2, t+2]$, where $t$ is the id of the shot. All frames and subtitles in the time window $T_t$ will be collected as the context.

\subsection{Experiment Settings}
Following the setting of TransReID\cite{he2021transreid}, all training images are resized to 256$\times$128 and augmented with random horizontal flipping, padding, random cropping and random erasing\cite{zhong2020random}. The batch size is set to 64 with 16 images per character ID for triplet loss. Our model is optimized by SGD with a 0.9 momentum setting. During the first 120 epochs, we only train the visual backbone with the initial learning rate of 0.08 and weight decay of 1e-4. During the last 20 epochs, DRWM and social context-aware GCN are added to train with visual search backbone parameters frozen. The initial learning rate is set to 1e-4 and weight decay is set to 1e-5. All experiments are conducted on two GeForce RTX 3090 with PyTorch and Torchvison toolboxes. 
 % \vspace{-0.3cm}

\subsection{Baselines}
% We totally selected 13 models as baselines. 
% \textbf{ResNet}\cite{he2016deep}, \textbf{ViT}\cite{dosovitskiy2020image} use ResNet/Transformer as the backbone to learn robust representation. \textbf{HACNN}\cite{li2018harmonious}, \textbf{MLFN}\cite{chang2018multi}, \textbf{ResNet-mid}\cite{yu2017devil}, \textbf{OSNet}\cite{zhou2019omni} catch multi branch information to learn characters' features. \textbf{OSNet-AIN}\cite{zhou2021learning} introduces instance normalization layers to adapt the styles across domains. \textbf{NFormer}\cite{wang2022nformer}, introduces a neighbor transformer network to eliminate outlier features. \textbf{TransReID}\cite{he2021transreid} introduces jigsaw patch module and side information embeddings to enhance the robustness of the model. \textbf{SCPS}\cite{li2021social} jointly using visual information and social relationships to assist character search. SCPS-L and SCPS-P represent linear and prudent label update strategies, respectively. \textbf{TP}\cite{li2021social}, which is proposed as baseline in \cite{li2021social}. it combines textual similarity with visual similarity to obtain the final measurement. TP-ResNet and TP-ViT represent TP with different visual backbones.

We compare our model with the following state-of-the-art methods to evaluate its performance:

\begin{itemize}
	\item \textbf{HACNN}\cite{li2018harmonious}, which contains two branches, one branch learns local features, and the other learns global features. At the same time, a attention selection module is adopted to learn a set of complementary attention maps to optimize the representation.
	
	\item \textbf{MLFN}\cite{chang2018multi}, which contains multiple factor modules and factor selection modules to catch the semantic information at different level, and dynamic select the information extracted from multiple factor modules to explain content of each image.
	
	\item \textbf{ResNet}\cite{he2016deep}, which is widely used in computer vision tasks\cite{ lin2019tsm, zhao2022me, zhao2022pedm} and can solve the learning degradation problem in traditional CNN-based networks.
	
	\item \textbf{ResNet-mid}\cite{yu2017devil}, which extracts the mid-level features and fuses them with the output of the last layer as the final representation, making itself capable to learn discriminative features at different semantic levels.
	
	\item \textbf{OSNet}\cite{zhou2019omni}, which contains a unified aggregation gate that can fuse multi-scale features caught by omni-scale residual blocks to learn homogeneous and heterogeneous information to improve its performance.
	
	\item \textbf{OSNet-AIN}\cite{zhou2021learning}, which introduces instance normalization layers to adapt the styles across domains on the basis of OSNet. In addition, an architecture search algorithm is performed to find the best locations for instance normalization layers.
	
	\item \textbf{NFormer}\cite{wang2022nformer}, which introduces a neighbor transformer network to eliminate outlier features and generate more reliable representations by modeling the relations between all the input images.
	
	\item \textbf{ViT}\cite{dosovitskiy2020image}, which splits the image into N patches, and then uses a linear projection function to map them into 1-D tokens. In this way, the image has converted into a sequence of tokens, which can be processed by transformer.
	
	\item \textbf{TransReID}\cite{he2021transreid}, which introduces jigsaw patch module and side information embeddings to strengthen the connection between different patches and learn domain style, enhancing the robustness of the model.
	
	\item \textbf{TP}\cite{li2021social}, which is proposed as baseline in \cite{li2021social}. Following their ideas, we use LSTM to extract text features and calculate the similarity, then combine textual and visual similarity to obtain the final measurement. TP-ResNet and TP-ViT represent as their visual backbones.
	
	\item \textbf{SCPS}\cite{li2021social}, which introduces a relation-aware framework to identify characters by using visual information and social relations revealed by multi-modal context. SCPS-L and SCPS-P represent as linear and prudent label update strategies, respectively.
\end{itemize}

\subsection{Overall Performance}
\begin{table}
\tabcolsep=0.11cm
        % \small{
	\begin{center}
		\caption{Overall performance on two datasets. TRM means Transformer. The best results are highlighted in bold.}
		\label{performance}
		\begin{tabular}{c|c|c|c|c|c|c|c}
			\toprule
			\multirow{2}{*}{Method} &\multirow{2}{*}{Backbone} & \multicolumn{3}{c}{Social-Bilibili} & \multicolumn{3}{c}{Social-MoviNet} \\
			 & & \multicolumn{1}{c}{mAP} & \multicolumn{1}{c}{mINP} & \multicolumn{1}{c}{R1} & \multicolumn{1}{c}{mAP} & \multicolumn{1}{c}{mINP} & \multicolumn{1}{c}{R1} \\
			\midrule
			\midrule
			  HACNN\cite{li2018harmonious} & CNN & 47.8 & 29.4 & 66.0 & 30.4 & 20.7 & 55.0 \\
			 MLFN\cite{chang2018multi} & CNN & 55.7 & 34.6 & 72.0 & 30.6 & 20.6 & 40.0\\
			 ResNet\cite{he2016deep} & CNN & 61.2 & 35.8 & 82.0 & 42.1 & 22.2 & 65.0 \\
			 ResNet-mid\cite{yu2017devil} & CNN & 63.8 & 35.3 & 82.0 & 45.0 & 22.0 & 75.0 \\
			 OSNet\cite{zhou2019omni} & CNN & 64.6 & 38.2 & 80.0 & 49.1 & 21.2 & 75.0 \\
			 OSNet-AIN\cite{zhou2021learning} & CNN & 65.1 & 39.7 & 86.0 & 45.3 & 20.9 & 80.0 \\

            TP-ResNet\cite{li2021social} & CNN & 59.5 & 36.2 & 78.0 & 44.8 & 21.7 & 75.0 \\
			\midrule
			\midrule
           NFormer\cite{wang2022nformer} & TRM & 63.8 & 36.3 & 80.0 & 37.8 & 20.7 & 65.0 \\
			 ViT\cite{dosovitskiy2020image} & TRM & 76.5  & 49.1  & 90.0 & 59.4  & 23.0  & 90.0 \\
			 TransReID\cite{he2021transreid} & TRM & 79.9 & 51.2 & 90.0 & 62.3 & \textbf{24.6} & 90.0 \\
			 
			 TP-ViT\cite{li2021social} & TRM & 68.1 & 40.3 & 86.0 & 59.2 & 24.1 & \textbf{95.0} \\
			 SCPS-L\cite{li2021social} & TRM & 80.9 & 52.6 & 90.0 & 57.9 & 23.9 & 80.0 \\
			 SCPS-P\cite{li2021social} & TRM & 81.0 & 53.1 & 90.0 & 58.9 & 24.1 & 80.0 \\
			 Ours & TRM & \textbf{82.8} & \textbf{56.3} & \textbf{94.0} & \textbf{63.6} & 24.5 & \textbf{95.0} \\
			\bottomrule
		\end{tabular}
	\end{center}
 % }
 \vspace{-0.6cm}
\end{table}
In Table \ref{performance}, we show the overall performance of our model and the baselines. Following the re-ID community, we select mAP and CMC-Rank@1(R1) as our evaluation metrics. In addition, because the video character search is used as the upstream task of idols summarization and the enthusiastic fans usually don't want to miss any shots of their idols, we need to evaluate the cost of searching for the last correct image. Therefore, we introduce mINP\cite{ye2021deep} to measure the ability to retrieve the hardest match of models.\par
Overall, our social context-aware method performs better than traditional methods. Taking TransReID as an example, on the Social-Bilibili dataset, our method reaches 82.8\%, 56.3\% and 94.0\%, surpassing TransReID 2.9\%, 5.1\% and 4.0\% on the terms of mAP, mINP and R1 respectively. This is because TransReID only uses the visual features and treats each character as an independent instance, while our method can integrate characters' relationships and co-occurrence to assist the character search. However, on the Social-MovieNet dataset, the mINP decreased by 0.1\% compared with TransReID. Because as the mAP of the visual search backbone decreases (the visual search task on the Social-MovieNet dataset is more challenging), the anchor node selection becomes less accurate, which makes it difficult to propagate features of the corresponding query node to the gallery node.\par
In addition, we compared the different methods of using social clues. Take SCPS-P as an example, on Social-Bilibili dataset, our model outperformed 1.8\%, 3.2\% and 4.0\% on mAP, mINP and R1 respectively. Compared with our method, SCPS doesn't consider the role of the prior probability of the relationship between the characters in different scenes, thus limiting the model performance. We can observe similar results on the Social-MovieNet dataset.\par
 % \vspace{-0.2cm}
\subsection{Ablation Study}
\begin{table}
	\begin{center}
		\caption{The performance of our method with different visual backbones on Social-Bilibili dataset}
		\label{ablationvbb1}
		\begin{tabular}{c|c|c|c|c}
			\toprule
			Method & Backbone & mAP & mINP & R1 \\
			\midrule
			\midrule
			ResNet & ResNet50 & 61.2 & 35.8 & 82.0 \\
			ResNet-mid & ResNet50 & 63.8 & 35.3 & 82.0 \\
			TransReID & DeiT-S/16 & 65.4 & 42.3 & 88.0 \\
			TransReID & DeiT-B/16 & 75.6 & 48.7 & 90.0 \\
			TransReID & ViT-S/16 & 64.8 & 39.2 & 88.0 \\
			TransReID & ViT-B/16 & 79.9 & 51.2 & 90.0 \\
			\midrule
			\midrule
			SoCoSearch & ResNet50 & 63.2 & 38.1 & 84.0 \\
			SoCoSearch & ResNet-mid & 63.9 & 37.7 & 82.0 \\
			SoCoSearch & DeiT-S/16 & 67.9 & 45.1 & 88.0 \\
			SoCoSearch & DeiT-B/16 & 78.5 & 52.9 & 94.0 \\
			SoCoSearch & ViT-S/16 & 67.3 & 41.0 & 86.0 \\
			SoCoSearch & ViT-B/16 & \textbf{82.8} & \textbf{56.3 }& \textbf{94.0} \\
			\bottomrule
		\end{tabular}
	\end{center}
 \vspace{-0.5cm}
\end{table}
\textbf{Different Backbones:}
In this section, we discuss the performance of our model with different visual backbones. The experimental results are shown in Table \ref{ablationvbb1}. We can observe that both CNN-based methods and transformer-based methods have improved after equipped social context-aware GCN on Social-Bilibili Dataset. In addition, the improvement of transformer-based methods is more significant than CNN-based methods'. Because SoCoSearch need to select anchors before building the social context graph, and the mAP of transformer-based methods are significantly higher than CNN-based methods', which makes this kind of method can select anchors more accurately. The correct anchors make social-context aware GCN more likely to aggregate the features of corresponding query characters to the gallery characters, thereby improving the model performance.

\textbf{Different Modalities:}
We also compare the performance of our model with different modal information. The experimental results are shown in Table \ref{ablationrbb1}. For models that neither uses visual nor textual context, it can be regarded as the TransReID, which only depends on the image features to match characters. Hence, it can not be well applied to the video characters search task. However, our model uses visual and textual information as clues to construct a social co-occurrence graph, which improves the similarity between the corresponding gallery and query nodes. It can be seen that when just using visual or textual information, the model is still effective, but the performance will be slightly decreased compared with the model with both modalities. Because a single modality may introduce bias, which leads to the DRWM being unable to allocate weights correctly, thus decreasing the model performance. Thus, it is necessary to integrate multimodal clues for better weights allocation.

\textbf{Different Balance Strategies:}
Finally, to verify our adaptive strategy's effectiveness, we compared different methods of balancing social features. Table \ref{ablationadp} shows three strategies to control social weight: mean, random, and adaptive weight. According to the mean and random strategy results, we can observe that inappropriate social weight will damage the original performance. However, our adaptive strategy is well-adapted to different scenes. It can adjust the social weight based on multimodal clues, making the model perform better.

\begin{table}
	\begin{center}
		\caption{The performance of our method with different modal information on Social-Bilibili dataset}
		\label{ablationrbb1}
		\begin{tabular}{c|c|c|c|c|c}
			\toprule
			Method & Visual & Textual & mAP & mINP & R1 \\
			\midrule
			\midrule
			 w/o VT & - & - & 79.9 & 51.2 & 90.0 \\
			 w/o V & - & Bert & 82.6 & 55.4 & 94.0 \\
			 w/o T & TSM & - & 82.7 & 56.2 & 94.0 \\
			all & TSM & Bert & \textbf{82.8} & \textbf{56.3} & \textbf{94.0} \\
			\bottomrule
		\end{tabular}
	\end{center}
 \vspace{-0.3cm}
\end{table}

\begin{table}
	\begin{center}
		\caption{The performance of our method with different balance strategies on Social-Bilibili dataset}
		\label{ablationadp}
		\begin{tabular}{c|c|c|c}
			\toprule
			Strategy & mAP & mINP & R1 \\
			\midrule
			\midrule
			Mean & 82.4 & 55.4 & 94.0 \\
			Random & 76.0 & 47.9 & 90.0 \\
			Adaptive(Ours) & 82.8(+0.4) & 56.3(+0.6) & 94.0(+0) \\
			\bottomrule
		\end{tabular}
	\end{center}
 \vspace{-0.3cm}
\end{table}

%% file: sections/conlusion.tex
\section{CONCLUSION}
In this paper, we proposed a social context-aware framework named SoCoSearch to deal with the video character search task. SoCoSearch can integrate prior knowledge and characters' co-occurrence to construct a social context graph. With the help of the graph, we can get a set of potential matching characters. And it employed a social context-aware GCN to pass the candidate node's features to the corresponding gallery characters to improve the similarity between the query-gallery pairs and complete the matching. Extensive experiments on the two real-world datasets demonstrated an improvement over several baselines, proving our method's superiority in the video character search task.